\pdfoutput=1
\documentclass[11pt,a4paper]{article}
\usepackage{jheppub}

\usepackage{graphicx}
\usepackage{dcolumn}
\usepackage{bm}
\usepackage{epsfig}
\usepackage{enumerate}

\usepackage[usenames,dvipsnames]{xcolor}

\usepackage{amsmath}
\usepackage{amssymb}

\newcommand{\RR}{{\rm R}}
\newcommand{\LL}{{\rm L}}

\notoc

\begin{document}

\preprint{ANL-HEP-PR-12-55}

\title{Dark Matter and enhanced $h \to \gamma \gamma$ rate from vector-like Leptons}

\author[a]{Aniket Joglekar,}
\emailAdd{aniket@uchicago.edu}
\author[b,c]{Pedro Schwaller}
\emailAdd{pschwaller@hep.anl.gov}
\author[a,b,d]{ and Carlos E. M. Wagner}
\emailAdd{cwagner@hep.anl.gov}

\affiliation[a]{Enrico Fermi Institute, University of Chicago, Chicago, IL 60637, U.S.A.}
\affiliation[b]{HEP Division, Argonne National Laboratory, 9700 Cass Ave., Argonne, IL 60439, U.S.A.}
\affiliation[c]{Physics Department, University of Illinois at Chicago, Chicago, IL 60607, U.S.A.}
\affiliation[d]{Kavli Institute for Cosmological Physics, University of Chicago, Chicago, IL 60637, U.S.A.}

\abstract{
In this paper, we study an extension of the standard model with a vector-like generation of leptons. This model provides a viable dark matter candidate and a possibility to enhance the Higgs decay rate into a pair of photons. We evaluate constraints from electroweak precision tests and from vacuum stability, and find that the latter provide an upper limit on the lepton Yukawa couplings. A large enhancement of the Higgs di-photon rate can therefore only be obtained when the mass of the lightest charged lepton is close to the LEP limit. 
The relic density constraint suggests a co-annihilation scenario with a small mass difference between the lightest charged and neutral leptons, which also weakens the LEP limit  on the lightest charged lepton mass and allows for larger Higgs di-photon decay rates. 
Cross sections for direct detection of the dark matter candidate are calculated, and prospects for detecting the new particles at the LHC are discussed briefly. 
}

\maketitle

\section{Introduction}\label{sec:into}

The discovery of a scalar resonance consistent with the SM Higgs boson~\cite{Gianotti:gia12,Incandela:inc12} marks a great success in the early run of the CERN Large Hadron Collider (LHC) and its associated experiments. It is now conceivable that electroweak symmetry is indeed broken by the Higgs mechanism, namely through the vacuum expectation value of a fundamental Higgs doublet. With a Higgs mass of about 125~GeV, this model is fully renormalizable and potentially stable up to very high scales. 

However some mysteries remain that suggest additional particles being present at the electroweak scale. First, the existence of dark matter has been established through astrophysical observations beyond reasonable doubt, however its nature and properties remain largely unknown. With all known matter content in the SM being related to the weak scale, it is at least a reasonable assumption that also the dark matter particle is connected to the electroweak scale. 

The second mystery is provided by the Higgs boson itself. While the existence of a new resonance has very recently been established, it remains to be seen whether its properties, in particular the production cross sections and decay branching ratios, agree with the very precise predictions from the SM. The decay channels that are most sensitive to new physics effects are the loop induced decays of the Higgs to pairs of photons or to a photon and a Z boson. And indeed the rate of di-photon events shows the largest and most consistent deviation from SM expectations, with both 2011~\cite{ATLAS:2012ad,Chatrchyan:2012tw} and 2012~\cite{Atlasnote:-2012-091,CMSnote:12-015} data from both the Atlas and CMS experiments suggesting an enhancement of the branching ratio of about 50\% (but still within SM expectations at the two sigma level\footnote{Very recent analyses of the combined ATLAS and CMS data sets suggest that the deviation of the di-photon signal from SM expectations is at the $2\sigma$-$2.5\sigma$ level~\cite{Giardino:2012dp,Espinosa:2012im,Low:2012rj,Ellis:2012hz,Carmi:2012in}.}). 

Motivated by the above, here we study a simple extension of the SM that provides a dark matter candidate and allows for an enhancement of the Higgs branching ratio to photon pairs. The model introduces a vector-like fourth generation of leptons, namely SU(2) doublets $\ell = (\ell_{\rm L}', \ell_{\rm R}'')$, charged  SU(2) singlets $e = (e_{\rm R}', e_{\rm L}'')$ and neutral singlets $\nu = (\nu'_{\rm R}, \nu''_{\rm L})$. In the limit where the vector-like masses vanish the particle content is that of a fourth generation of leptons (indicated by a single prime) and an exact copy with opposite chirality (double primed). 

The remainder of the paper is organized as follows. In the next section, the particle content of the model is introduced and the mixing in the charged and neutral sector is discussed. Electroweak precision tests and other constraints on the model are presented in Sec.~\ref{sec:EWP} while the modified Higgs boson properties are analyzed in detail in Sec.~\ref{sec:higgs}. In Sec.~\ref{sec:rge} constraints on the magnitude of the Yukawa couplings from vacuum stability considerations are derived. Dark matter properties are explored in Sec.~\ref{sec:dm} before we conclude in Sec.~\ref{sec:conclusions}. Formulas for the electroweak precision observables are presented in  detail in the appendix.

\section{The Model}\label{sec:model}

The model introduces one additional family of standard model (SM) like leptons consisting of a left-handed SU(2) doublet $\ell_{\rm L}'$ and the corresponding charged and neutral singlets $e_{\rm R}'$ and $\nu_{\rm R}'$ to allow mass generation through the Higgs mechanism. In addition one mirror family of leptons is introduced with identical quantum numbers, but opposite chirality. This makes the model manifestly anomaly free, since the fields of opposite chirality combine into vector-like multiplets. The full particle content introduced in this model is displayed in Tab.~\ref{tab:fields}. 

The Lagrangian density of the model is given by 
\begin{align}
	{\cal L} & = -m_\ell \bar{\ell}'_\LL  {\ell }''_\RR - m_e \bar{e}''_\LL {e}'_\RR - m_\nu \bar{\nu}_\LL'' {\nu}_\RR' - \frac{1}{2}m' \overline{{\nu'_\RR}^c} \nu'_\RR - \frac{1}{2}m'' \overline{{\nu''_\LL}^c} \nu ''_\LL\notag \\
	&\,\,\,\, -Y_c' (\bar{\ell}'_\LL H ) {e}_\RR' - Y_n' (\bar{\ell}'_\LL \tau H^\dagger)  {\nu}_\RR' - Y_c'' (\bar{\ell}''_\RR H ) {e}_\LL'' - Y_n'' (\bar{\ell}''_\RR \tau H^\dagger)  {\nu}_\LL'' + \rm{h.c.} 
\end{align}
where the standard kinetic terms are omitted. 
We have written down all mass terms and couplings that are allowed by gauge invariance, including Majorana masses for the neutral singlet neutrinos. Couplings that would mix the new leptons with SM leptons are set to zero for the remainder of this paper. Vanishing of these couplings can be guaranteed by imposing a parity symmetry on the new sector, similar to~\cite{Lee:2012xn}.  For the remainder of this paper, we will always assume that such a symmetry is present.

In terms of vector-like multiplets, the particle content can be written as $\ell = (\ell_{\rm L}', \ell_{\rm R}'')$ and $e = (e_{\rm L}'',e_{\rm R}')$.
\begin{table}
\begin{tabular}{| l |c|c|c|c|c|c|}
\hline
Name & $\ell_{{\rm L}}'$ & $e_{{\rm R}}'$ & $\nu_{\rm{R}}'$ & $\ell_{{\rm R} }''$ & $e_{\rm L}''$ & $\nu_{\rm L}'' $ \\ \hline
Quantum Numbers & (1,2,-1/2) & (1,1,-1) & (1,1,0) & (1,2,-1/2) & (1,1,-1) & (1,1,0) \\ \hline
\end{tabular}
\caption{Particle content of the model with ${\rm SU(3)} \times {\rm SU(2)} \times {\rm U(1)}$ quantum numbers, in the chirality basis. \label{tab:fields}}
\end{table}
To avoid confusion between the flavor and mass eigenstates, the notational conventions are chosen as follows:
\begin{itemize}
	\item Flavor basis: Lower case particle names $\nu$, $e$ and mass parameters $m_x$.
	\item Mass basis: Upper case particle names $N_{1,2,3,4}$ and $E_{1,2}$, and corresponding masses $M_{N_i}$, $M_{E_i}$.
\end{itemize}

There are two interesting limits of the above Lagrangian that can be realized by imposing a discrete symmetry on some of the new fields. First, one can impose a  second parity symmetry under which the new SU(2) singlet fields are odd, while all other fields are even. This symmetry forbids Yukawa couplings between the new fields such that $Y_c' = Y_c'' = Y_n' = Y_n'' = 0$, and the masses for all fields come exclusively from explicit vector-like mass terms in the Lagrangian, and are not affected by electroweak symmetry breaking. In this limit the new fields can be decoupled from the SM without any observable low energy effects. While we will not  discuss this possibility in detail, the decoupling limit is useful for checking the calculation of the electroweak S and T parameters. 

Alternatively, one can impose an additional discrete symmetry under which the mirror leptons are odd, such that the explicit mass terms $m_\ell$, $m_e$ and $m_\nu$ are forbidden. In this limit, the mirror sector can not mix with SM leptons, however the new SM like leptons will behave like an ordinary fourth lepton family, with corresponding experimental signatures and limits. A model with a similar leptonic sector was also considered in~\cite{Ishiwata:2011hr,Arnold:2012fm}. 
For the remainder of this paper we will focus on the most general case, where only the parity symmetry that forbids mixing with ordinary SM leptons is present.

\

In the limit where the explicit mass terms vanish, the
spectrum of the model can easily be derived from the Lagrangian. After electroweak symmetry breaking, there are two charged leptons with masses $Y_c' v$ and $Y_c''v$, where $v=174$~GeV is the Higgs vacuum expectation value (VEV). In the neutral sector the two massive neutrino states are further split when the Majorana masses are nonzero, such that there are four neutrinos with masses
\begin{align}
M_{N_{1,2}}&=\sqrt{\frac{m'^2}{4}+Y'^2v^2}\pm\frac{m'}{2}\,,\\
M_{N_{3,4}}&=\sqrt{\frac{m''^2}{4}+Y''^2v^2}\pm\frac{m'}{2}\,.
\end{align}

When all masses and Yukawa couplings are non-zero, the spectrum is slightly more complicated, since now there is mixing between the ordinary and the mirror leptons. In the charged sector, the $2\times 2$ dimensional mass matrix ${\cal M}_c$ is defined as

\begin{align}
	{\cal L} \supset 
	\begin{pmatrix} \overline{{e}_\LL' }&  \overline{{e}_\LL''} \end{pmatrix} {\cal M}_c \begin{pmatrix} {e}_\RR' \\ {e}_\RR'' \end{pmatrix} + {\rm h.c.}
	\qquad \text{with} \qquad {\cal M}_c = 
	\begin{pmatrix} Y_c' v & m_\ell \\ m_e & Y_c'' v \end{pmatrix}.
\end{align}
In the neutrino sector all four states are mixed with a symmetric mass matrix ${\cal M}_n$ given by
\begin{align}
	\frac{1}{2}\begin{pmatrix}
         \overline{\nu'_{\rm L}} & \overline{{\nu'_{\rm R}}^c} & \overline{{\nu''_{\rm R}}^c} & \overline{\nu''_{\rm L}}\end{pmatrix}{\cal M}_n\begin{pmatrix}
	{\nu'_{\rm L}}^c\\        
	\nu'_{\rm R}\\ 
        \nu''_{\rm R}\\
        {\nu''_{\rm L}}^c
	\end{pmatrix}+ {\rm h.c.}
	\qquad \text{with} \qquad {\cal M}_n = 
	\begin{pmatrix}
	0 & Y'_\nu v & m_\ell & 0 \\
	Y'_\nu v & m' & 0 & m_\nu\\
        m_\ell & 0 & 0 & Y''_\nu v\\
        0 & m_\nu & Y''_\nu v & m''
	\end{pmatrix}.
\end{align}
The mass matrices can be diagonalized as follows:
\begin{align}
{\cal M}_{cD} &= U_L^\dagger {\cal M}_c U_R \,,\\
{\cal M}_{nD} & = V^T {\cal M}_n V \,,
\end{align}
where ${\cal M}_{cD}$ and ${\cal M}_{nD}$ are diagonal positive semi-definite matrices and $U_L$, $U_R$ and $V$ are unitary. 
We denote the mass eigenstates as $E_1$ and $E_2$ in the charged sector and $N_{1-4}$ in the neutral sector with corresponding masses $M_{E_i}$ and $M_{N_i}$, ordered such that $M_i<M_j$ when $i<j$.

\section{Electroweak Precision Tests and Direct Limits}\label{sec:EWP}

In the limit where new particles are heavier than $M_Z/2$, the effects on electroweak precision observables can be very well estimated by calculating the contributions to the so called S and T parameters~\cite{Peskin:1991sw}. 

We have calculated the contributions to the oblique electroweak parameters S and T for the most general allowed mass matrices ${\cal M}_c$ and ${\cal M}_n$. The formulas are given in the appendix. To our knowledge, these results are not available in the literature. Formulas for limiting cases are however available, and have been used to verify our results. In particular, the case of a chiral generation of leptons that mixes with the SM was treated in detail in~\cite{Eberhardt:2010bm}, while some results for mixed vector-like fermions can be found in~\cite{Cynolter:2008ea}.

For a Higgs mass of $m_H=125$ GeV, the 95\% CL allowed ranges for the S and T parameters for $U=0$ are~\cite{PDG}
\begin{align}
S_{\text{best fit}}&=0.04\,,\quad\quad\quad\sigma_\text{S}=0.09\,,\notag\\
T_{\text{best fit}}&=0.07\,,\quad\quad\quad\sigma_\text{T}=0.08\,,\notag\\
\rho_{\text{corr}}&=0.88\,,
\end{align}
where $\sigma_{S,T}$ denotes the $1\sigma$ error and $\rho_{\rm corr}$ denotes the correlation between the two quantities. We discuss some regions of parameter space that are relevant for the remainder of this paper in the following. 

Obviously when all Yukawa couplings are zero the new particles decouple from the electroweak sector and $\Delta S = \Delta T = 0$, as long as the explicit mass terms are large enough for the S-T formalism to be a valid approximation.

\begin{figure}
\center
\includegraphics[width=.45\textwidth]{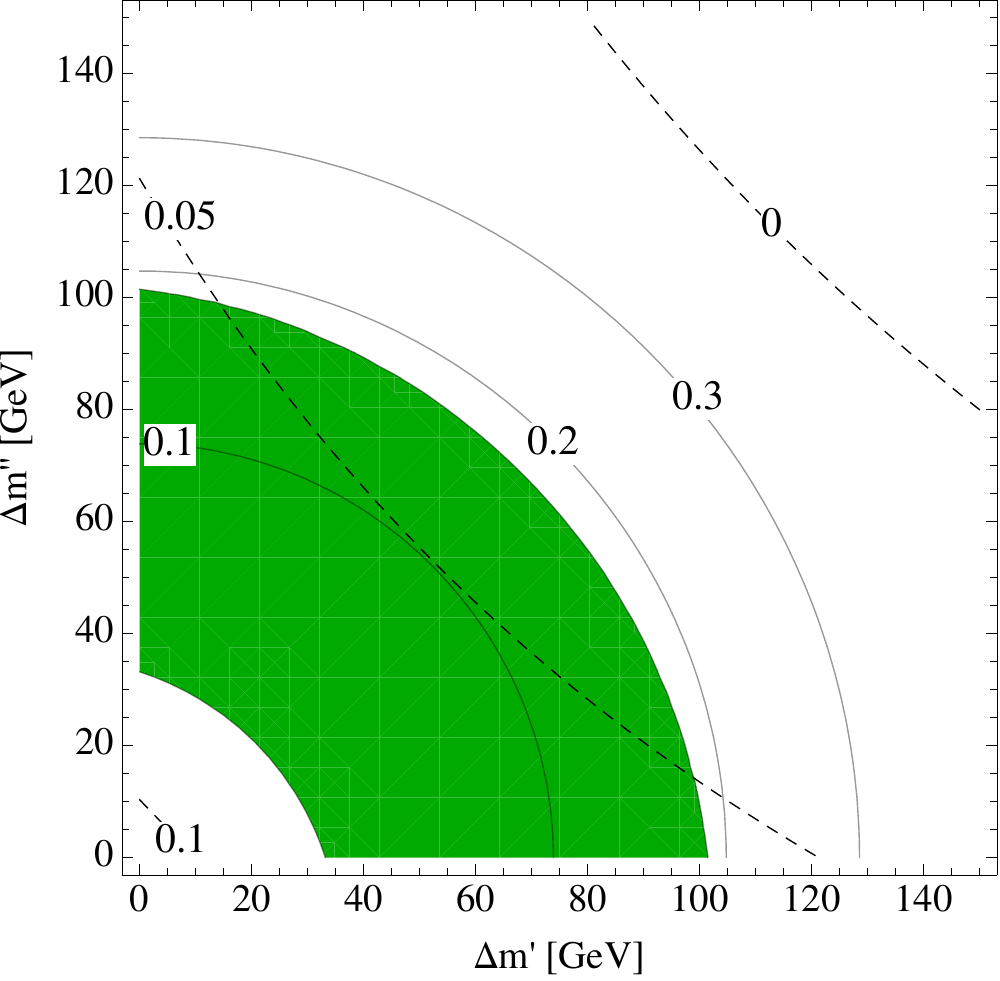}\hspace*{1cm}
\includegraphics[width=.45\textwidth]{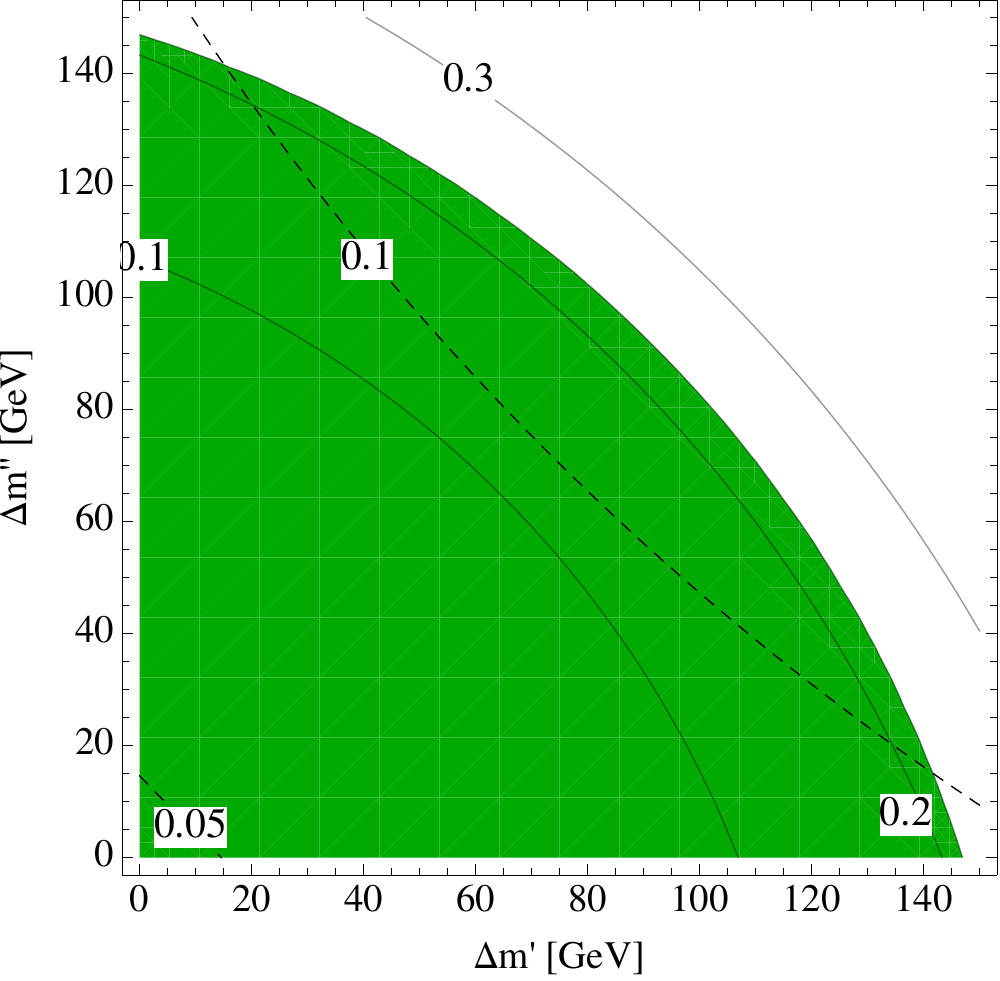}
\caption{Decoupling of effects on electroweak precision observables. The green (grey) shaded region shows the allowed parameter space as function of the mass splitting parameters $\Delta m' = (Y_c' - Y_n') v$ and $\Delta m'' = (Y_c'' - Y_n'')v$. Also shown are contours of constant $\Delta T$ (solid) and $\Delta S$ (dashed). On the left the vector-like masses are all vanishing, while the right plot is for $m_\ell = m_e = m_\nu = 400$~GeV. 
}
 \label{fig:st1}
\end{figure}

The opposite limit, when only the Yukawa couplings are nonzero, is also well known. When the splitting between the members of the doublet goes to zero, i.e. $Y_n' = Y_c'$ and $Y_n'' = Y_c''$, there is a positive contribution of to the S parameter of $\Delta S \approx 0.11$, while $\Delta T=0$, which lies outside of the 95\% CL region. Going away from the degenerate limit, $\Delta T$ grows proportional to the square of the mass difference of the members of each doublet, while $\Delta S$ receives corrections that are logarithmic in the mass difference. Due to the correlation between $S$ and $T$ this leads to a preference for a small splitting between the charged and neutral members of the doublets, with an upper limit on the splitting of about $\Delta m\sim 100$~GeV if only one doublet is non-degenerate, as illustrated in the left plot of Fig~\ref{fig:st1}. 

The degenerate case where the masses within one doublet are equal is also excluded, since here the positive shift in the $S$ parameter is not compensated by a small positive shift in $T$. This region
 can be brought into agreement with the  $S$ parameter by increasing the Majorana masses, since they tend to reduce $\Delta S$. 

Later we will be interested in scenarios with order one charged lepton Yukawa couplings, while keeping the neutral Yukawas small. In the absence of explicit vector-like masses such a scenario would be excluded by electroweak constraints. However here we can make use of the fact that the new sector decouples from the SM when the explicit mass terms $m_\ell$, $m_e$ and $m_\nu$ are increased while keeping the Yukawa couplings fixed. We therefore also expect that when $m_{\rm vector} \gtrsim Y_c v$, the contributions to the $S$ and $T$ parameters should be reduced, since we are transitioning to the decoupling regime. This effect is shown in the right plot of Fig.~\ref{fig:st1} where in addition to the nonzero Yukawa couplings the explicit mass terms are set to $m_\ell = m_e = m_\nu = 400$~GeV. For similar splittings between the charged and neutral Yukawa couplings, the contribution to the T parameter is suppressed compared to the chiral limit, such that a much larger region of parameter spaces is in agreement with precision constraints. 

\begin{figure}
\center
\includegraphics[width=.7\textwidth]{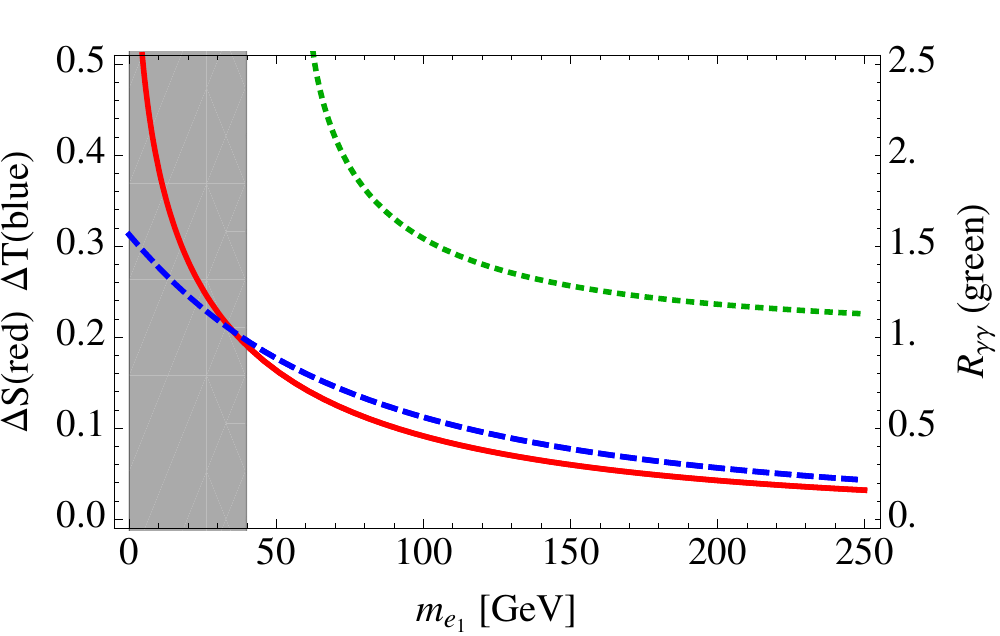}
\caption{Decoupling of $S$ (red, solid) and $T$ (blue, dashed) with increasing vector masses $m_\ell = m_e$, for $Y_c'=Y_c'' =0.8$, $Y_n' = Y_n''=0$ and $M' =M''=100$~GeV, as function of the lightest charged lepton mass $m_{e_1} = m_\ell - Y_c v$. Also shown is the ratio $R_{\gamma \gamma}$ (green, dotted) that is introduced in Sec.~\ref{sec:higgs}.}
 \label{fig:st2}
\end{figure}

In Fig.~\ref{fig:st2} the decoupling behavior is shown for $S$ and $T$ parameters separately. Here we choose a parameter region that will be relevant later, namely order one charged Yukawas with vanishing or small neutrino Yukawa couplings. To ensure that all neutrinos have sufficiently large masses the Majorana mass terms are set to $M'=M'' = 100$~GeV. Further we choose $Y_c' = Y_c'' = 0.8$ and $m_\ell = m_e$, so that the mass of the lightest charged lepton follows the simple relation 
\begin{align}
	M_{E_1} = | m_\ell - Y_c v| \,.
\end{align}
Only the region where $m_\ell > Y_c v$ is considered here. It can be seen in Fig.~\ref{fig:st2} that the S and T parameters decouple very fast in this regime. Sizable contributions only happen for charged lepton masses below $100$~GeV, however even this region is still in agreement with the experimental limits due to the positive correlation between $S$ and $T$. 

\

In addition to electroweak precision tests, there are also limits on the masses of new charged and neutral particles from the LEP experiments. 
The measurement of the width of the Z-boson at LEP constrains number of active neutrinos below 45 GeV to 3, which implies a lower bound of $M_{N} > 45$~GeV for a Dirac and $M_{N}>39$~GeV for an additional Majorana neutrino~\cite{PDG}. Note however that these limits can be partially evaded when the coupling of the lightest neutrino to the Z boson is suppressed. 

Direct searches at LEP also impose lower bounds on masses of the new charged leptons. If we assume that the lightest new neutrino state $N_1$ is stable on collider time-scales, the relevant decay is $E_1 \to W^{(*)} N_1 \to \ell + {E\!\!\!/}$, where $\ell$ denotes a SM lepton coming from the $W$-boson decay. Production of $E_1^+ E_1^-$ pairs at LEP with subsequent decay to SM leptons and missing energy puts a limit of $M_{E_1} > 101.9$~GeV on the mass of the lightest new charged lepton, provided that $\Delta M_1 = M_{E_1} - M_{N_1} > 15$~GeV.
When the mass difference between $E_1$ and $N_1$ is below this value, the resulting lepton can be very soft and therefore unobservable in the LEP detectors. In that case a weaker limit of $M_{E_1}>81.5$~GeV applies for $\Delta M_1>8.4$~GeV and $M_{E_1} > 63.5$ for $\Delta M_1 > 7$~GeV. No limits are available for smaller $\Delta M_1$, however limits from chargino searches~\cite{PDG} are available even for lower mass splittings, so that charged lepton masses below $80$~GeV should be treated with care. 

In the absence of mixing between the new leptons and the SM leptons, there are no constraints from lepton flavor and lepton number violation, such that the electroweak precision tests and the direct search limits from LEP are the most important constraints on the parameter space for our model.

\section{Higgs Properties}\label{sec:higgs}

The couplings of the mass eigenstates to the Higgs boson are then given by the diagonal entries of the rotated Yukawa coupling matrix $C_h = U_L^\dagger Y_c U_R$:
\begin{align}
	C_{h11} & =  Y_c' U_{L11}^* U_{R11} + Y_c'' U_{L21}^* U_{R21}\,,\\
	C_{h22} & = Y_c' U_{L12}^* U_{R12} + Y_c'' U_{L22}^* U_{R22}\,.
\end{align}
The new charged leptons, $E_1$ and $E_2$ contribute to the Higgs decay to photons at the one loop level. 
In the mass basis, the contributions to the decay are given by
\begin{align}\label{eqn:hgg_loop}
	\Gamma_{h \to \gamma\gamma} \propto \left| A_1(\tau_w) + \frac{4}{3}A_{1/2}(\tau_t) + \frac{C_{h11} v}{M_{E_1}}A_{1/2}(\tau_{E_1}) + \frac{C_{h22}v}{M_{E_2}}A_{1/2}(\tau_{E_2}) \right|^2\,,
\end{align}
where the loop functions for spin 1/2 and spin 1 particles are given by~\cite{Djouadi:2005gi}
\begin{align}
A_{1/2} (\tau) & = \frac{2\left(\tau+\left(\tau-1\right)f\left(\tau\right)\right)}{\tau^2} & \text{for spin  }\; \frac{1}{2} \,,\\
A_1(\tau) &=-2-\frac{3}{\tau}-\frac{3\left(2\tau-1\right)f\left(\tau \right) }{\tau^2 } & \text{for spin  } \;1\,,
\end{align}
with
\begin{align}
\tau_x&=\frac{m_h^2}{4m_x^2}\\
f\left(\tau\right) &= \begin{cases} \arcsin^2\sqrt{\tau}  & \text{for } \tau\leq1\,,\\
\displaystyle
-\frac{1}{4}\left(-i\pi+\log\left[\frac{1+\sqrt{1-\tau^{-1} }}{1-\sqrt{1-\tau^{-1}}}\right]\right)^2\quad \quad &\text{for } \tau>1\,,
\end{cases}
\end{align}
and $m_x$ is the mass of the particle running in the loop. 
It is instructive to consider the asymptotic values of the loop functions for $\tau_x\ll 1$, i.e. when the new particles are much heavier than the Higgs boson. First, the SM contribution with $m_h=125$~GeV is $A_1(\tau_w)=-8.3$ from the W-boson loop and $\frac 4 3 A_{1/2}(\tau_t) = 1.8$ from the top quark loop. Asymptotically $A_{1/2}(\tau \to 0) = 4/3$, while $A_{1/2}(\tau) > 4/3$ for $0<\tau<1$.

Since the new leptons don't affect the Higgs production channels, the effect on the di-photon search channel at the LHC is fully described by the ratio
\begin{align}\label{eqn:rgamma}
	R_{\gamma \gamma} =\frac{\sigma(pp \to h)}{\sigma_{\rm SM}(pp \to h)} \frac{\Gamma(h \to \gamma\gamma) }{\Gamma(h\to \gamma\gamma)_{\rm SM}} = \frac{\Gamma(h \to \gamma\gamma) }{\Gamma(h\to \gamma\gamma)_{\rm SM}}\,.
\end{align}
In the limit of vanishing Dirac mass terms $m_\ell$, $m_e$,
 the pre factors $c_{hii} v/m_i$ in~(\ref{eqn:hgg_loop}) go to one. It then follows that there is destructive interference between the dominant $W$ boson contribution and the charged lepton loops. In this limit, we find a large suppression of the di-photon rate, $R_{\gamma\gamma}<0.35$, across the range of allowed Yukawa couplings.

\begin{figure}
\center
\includegraphics{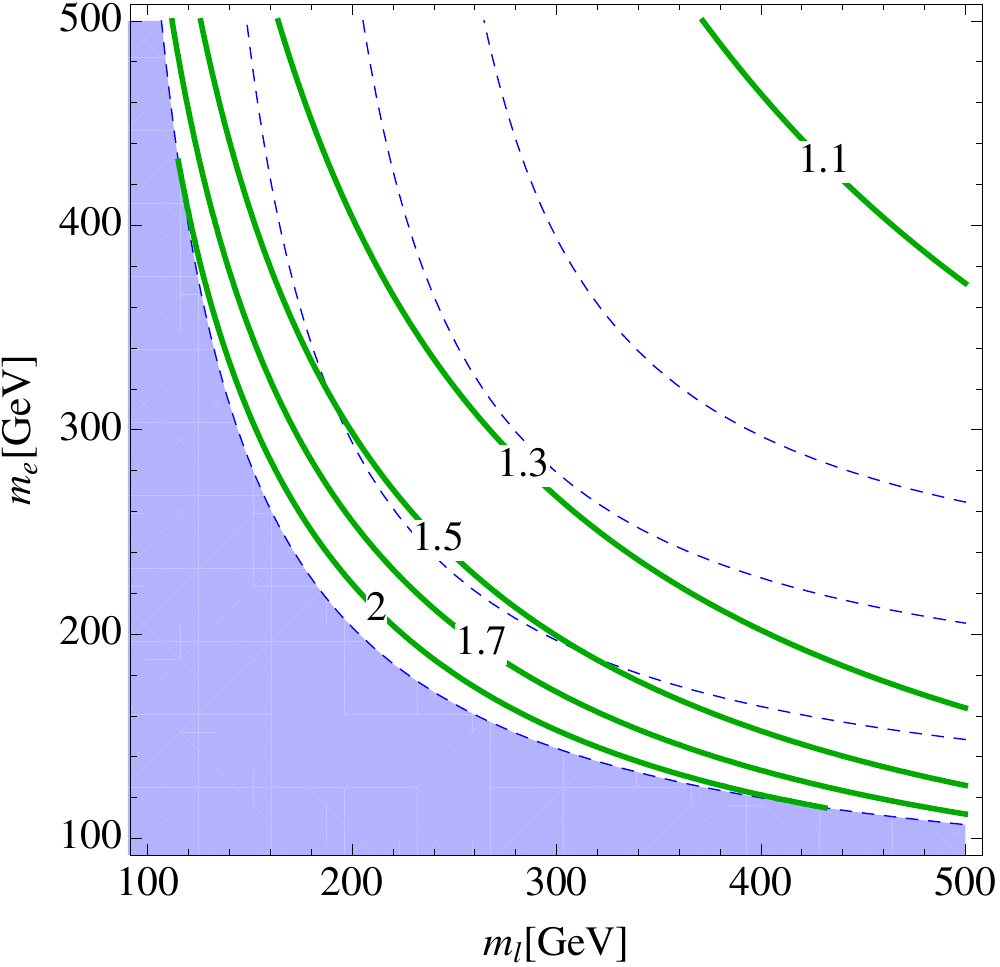}
\caption{Contours of constant $R_{\gamma\gamma}$ (green, solid) for $Y_c'=Y_c''=0.8$ as a function of the explicit mass terms $m_\ell$ and $m_e$. The blue (grey) shaded region indicates a mass for the lightest charged lepton below $62.5$~GeV, while the blue, dashed contours indicate a charged lepton mass $m_{E_1}$ of 62.5, 100, 150, and 200~GeV.}
 \label{fig:rgg1}
\end{figure}

To better understand the conditions for obtaining an enhancement in $R_{\gamma \gamma}$, we note that in the asymptotic limit the charged lepton contribution to the amplitude can be written as
\begin{align}\label{eqn:higgs_low_energy}
	A_{1/2}(0) \sum_{i} \frac{C_{hii} v}{m_i} = \frac{4}{3} v\frac{d}{dv} \log\det({\cal M})  =\frac{4}{3} \frac{ 2Y_c' Y_c'' v^2}{Y_c' Y_c'' v^2 - m_\ell m_e} \,,
\end{align}
where the first equality is a consequence of the Higgs low energy theorem~\cite{Ellis:1975ap,Shifman:1979eb,Falkowski:2007hz,Carena:2012xa}. Here it can also easily be understood by noting that $C_h = \partial {\cal M}/ \partial v$ and  remembering that ${\rm tr} (\log {\cal M}) = \log( \det {\cal M})$. 
Constructive interference occurs when~\ref{eqn:higgs_low_energy} becomes negative (and non-zero). For this to happen, both Yukawa couplings and both mass terms must be non-zero, and $m_\ell m_e > Y_c' Y_c '' v^2$ must be satisfied.

For order one Yukawa couplings, it is possible to obtain an enhancement of $\Gamma(h\to \gamma \gamma)$ up to 50\% compared to the SM prediction. This is illustrated in Fig.~\ref{fig:rgg1}, where we choose $Y_c' = Y_c'' =0.8$ and vary the Dirac mass terms to determine the regions where the di-photon rate is enhanced or suppressed compared to the SM. 

Larger enhancements can be obtained in two different ways. First, one could increase the charged Yukawa couplings to values above one to get ratios $R_{\gamma \gamma}$ of two or larger. However as we will see in the next section, such large Yukawas destabilize the Higgs potential below the TeV scale, such that an extension of the model would be required to realize such a scenario. Second, it would be possible to lower the mass of the lightest charged lepton below the LEP limit, and at the same time tune the mass of the lightest neutral state such that the decay $E_1 \to W^{(*)} N_1$ produces a very soft lepton that escapes detection. In that case an enhancement of up to 70\% can be obtained without further increasing the Yukawa couplings. 

\begin{figure}
\center
\includegraphics[width=0.7\textwidth]{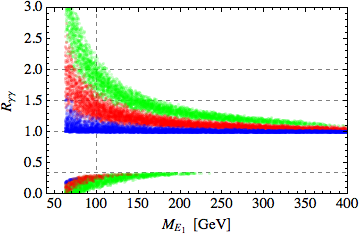}
\caption{The ratio $R_{\gamma\gamma}$ for random values of $m_\ell \in [0,600]$~GeV, $m_e \in [0,600]$~GeV and $Y_c',\; Y_c'' \in [0,0.5]$ (blue), $Y_c',\; Y_c'' \in [0.5,1.0]$ (red), and $Y_c',\; Y_c'' \in [1.0,1.5]$ (green). On the x-axis we show the mass of the lightest charged lepton. }
 \label{fig:random}
\end{figure}

To get a better idea about attainable values of $R_{\gamma \gamma}$ we have performed parameter scans with random choices for the mass terms and Yukawa couplings. In particular, we have varied $m_\ell$ and $m_e$ between $0$~GeV and $600$~GeV and the Yukawa couplings $Y_c'$ and $Y_c''$ in the range $0-1.5$. In Fig.~\ref{fig:random} we show the correlation between the rate $R_{\gamma \gamma}$ and the mass of the lightest charged lepton. Obviously the maximal value of $R_{\gamma \gamma}$ increases with the Yukawa couplings, however more importantly, it depends crucially on the mass of the lightest charged lepton.
The di-photon rate increases for lower values of the lightest charged lepton
mass, which is bounded from below by the LEP limits discussed in Sec.~\ref{sec:EWP}. 

Another interesting observation is that a small suppression of $R_{\gamma \gamma}$ can not be obtained in our model - either $R_{\gamma \gamma} > 1$ or $R_{\gamma \gamma} \lesssim 0.3$.
The latter possibility, which includes the limit of $m_\ell = m_e = 0$,
 is excluded by the recent results from the LHC experiments at the $3\sigma$ level. Note however that it is possible to populate the region  $0.3<R_{\gamma \gamma}<1$ when one of the explicit mass parameters is chosen to be negative. Since such a choice of parameters always leads to a suppression of $R_{\gamma\gamma}$ below one, we have not considered it further in the present work. 
 
 \
 
 Experimentally the ratio $R_{\gamma \gamma}$ is not directly observable, since the strength of the photon signal also depends on the Higgs production cross section. Defining $R_{ZZ}$ in analogy with~(\ref{eqn:rgamma}), one should instead consider the ratio $R_{\gamma \gamma}/R_{ZZ}$ to establish a deviation of the Higgs branching ratios from the SM predictions. The prediction of our model is that the Higgs production cross section and the decays to vector bosons and SM fermions are not modified, and therefore $R_{\gamma \gamma}/R_{ZZ} = R_{\gamma\gamma}$, in line with the current experimental data that shows no significant deviation of the signal rate in the $ZZ^*$ channel. 
 
A number of papers devoted to modifications of $R_{\gamma \gamma}/R_{ZZ}$ through effects of new particles have appeared recently. General requirements on new physics to obtain a positive contribution are discussed in~\cite{Carena:2012xa}. Analyses for specific models can be found in~\cite{Carena:2011aa,Cao:2012fz,Carena:2012gp,Benbrik:2012rm,An:2012vp,Buckley:2012em} for supersymmetric models
\footnote{In supersymmetric models it is also possible to obtain an enhanced $R_{\gamma\gamma}$ by suppressing the Higgs decay to bottom quarks~\cite{Ellwanger:2011aa,Gunion:2012gc,Barger:2012ky}. Such scenarios lead to $R_{\gamma\gamma}/R_{ZZ}\approx 1$, which is a clear distinction from the predictions of our model.}, 
in~\cite{Akeroyd:2012ms,Batell:2011pz,Arhrib:2011vc,Arhrib:2012ia,Wang:2012zv,Chang:2012ta,Chiang:2012qz} for scalar extensions of the SM and in~\cite{Goertz:2011hj,Bonne:2012im} for models with additional fermions. 

Since in all models the modifications are through loop effects, one possibility to distinguish between different models is to compare $R_{\gamma \gamma}$ with $R_{Z\gamma}$ since this Higgs decay is also loop induced~\cite{Carena:2012xa} and eventually will be observable at the LHC~\cite{Gainer:2011aa}. Since the W-boson loop fully dominates the Higgs decay to $Z\gamma$, new fermions will in general have a smaller effect in this decay. Indeed we have calculated $R_{Z\gamma}$ in our model and find deviations from the SM prediction of at most 5\% even in parameter regions where $R_{\gamma \gamma}$ is enhanced by 50\% or more.

\section{UV stability and Unification}\label{sec:rge}

\begin{figure}
\center
\includegraphics{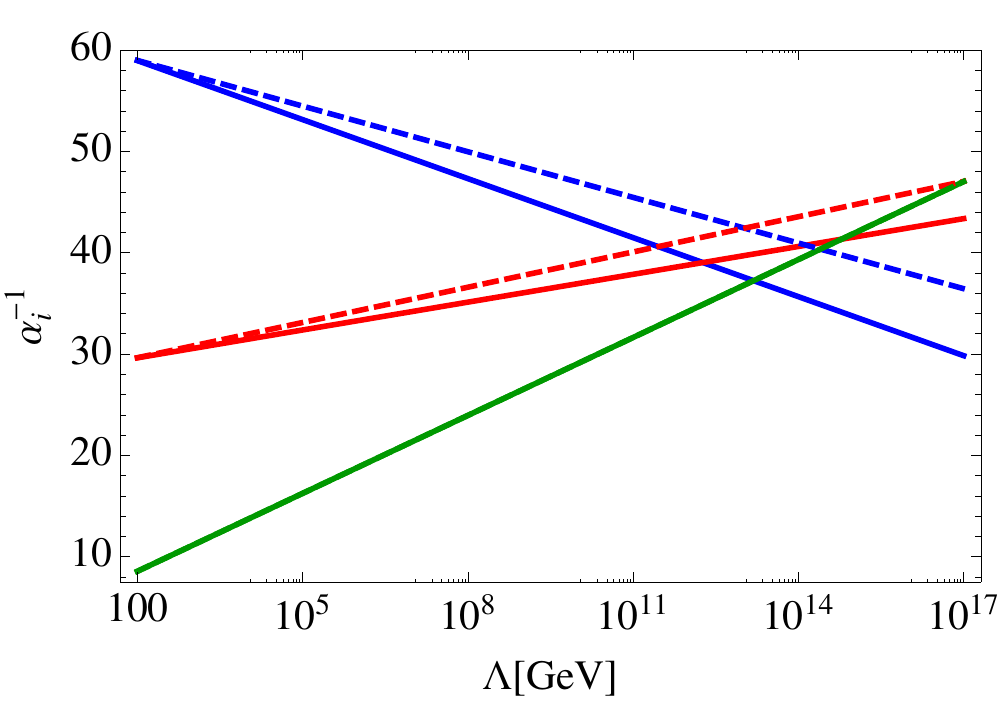}
\caption{One loop running of the gauge couplings (solid lines) compared with SM running (dashed lines).  From top to bottom $\alpha_1^{-1}$ (blue), $\alpha_2^{-1}$ (red) and $\alpha_3^{-1}$ (green) are shown, and both green curves lie on top of each other. }
 \label{fig:unification}
\end{figure}

The addition of particles charged under the electroweak gauge group and with nonzero Yukawa couplings will affect the running of gauge couplings and of the Higgs quartic coupling. At the one loop level, the RGE for the gauge couplings can be solved analytically and the running is given by
\begin{align}
	\alpha^{-1}_i (\Lambda) = \alpha^{-1}_i (m_Z) - \frac{b_i}{2\pi} \log\left(\frac{\Lambda}{m_Z}\right),
\end{align}
where $\alpha_i = g_i^2/(2 \pi)$ and $b_1 = \frac{53}{10}$, $b_2 = \frac{-15}{6}$ and $b_3 = -7$, while the corresponding SM values are $\frac{41}{10}$, $\frac{-16}{9}$ and $-7$. The evolution for scales up to $\Lambda = 10^{17}~{\rm GeV}$ is shown in Fig.~\ref{fig:unification}. Compared with the standard model, the curves for $\alpha^{-1}_{1,2}$ are bent downwards, such that unification is slightly improved and happens at a lower scale, around $\Lambda = 10^{14}~$GeV.  

For the running of the Yukawa couplings, we neglect the light quarks and leptons, and also assume that the neutrino Yukawa couplings are negligible compared to the charged lepton Yukawa couplings. In this limit, the beta functions for the top Yukawa $Y_t$ and the charged lepton Yukawas $Y_c'$ and $Y_c''$ are given by~\cite{Machacek:1983tz,Lindner:1985uk}
\begin{align}
	\beta_t & = \frac{1}{16 \pi^2}
  Y_t\left( \frac{3}{2}Y_t^2 + (3 Y_t^2 + {Y_c'}^2 + {Y_c''}^2) - 
    4 \pi \left(\frac{17}{12} \alpha_1 + \frac{9}{4}\alpha_2 + 8\alpha_3\right)\right), \\
	\beta_{Y_\ell} & = \frac{1}{16 \pi^2}
  Y_\ell\left( \frac{3}{2}Y_\ell^2 + (3 Y_t^2 + {Y_c'}^2 + {Y_c''}^2) - 
    4 \pi \left(\frac{15}{4} \alpha_1 + \frac{9}{4}\alpha_2 \right)\right),
\end{align}
where $Y_\ell = Y_c', \, Y_c''$. Finally the beta function for the Higgs quartic coupling, $\beta_\lambda$ is 
\begin{align}
	\beta_\lambda &=  \frac{1}{16 \pi^2} \left (12 \lambda^2 - 
   4 \pi \lambda (3 \alpha_1 + 9 \alpha_2) + (4\pi)^2 \left(\frac{3}{4}\alpha_1^2 + 
      \frac{3}{2} \alpha_1\alpha_2 + \frac{9}{4} \alpha_2^2\right) + \lambda  G - H\right),\\
	G & = 4\left( 3 Y_t^2 + Y_c'^2 + Y_c''^2\right),\qquad 
	H  = 4\left( 3 Y_t^4 + Y_c'^4 + Y_c''^4\right).
\end{align}
Thresholds are treated naively by setting the top quark and lepton Yukawa couplings to zero below the respective particle masses. 

It is well known that for a light Higgs the quartic coupling at the weak scale is small, such that the RGE evolution is dominated by the large top Yukawa in the SM, and decreases as the RGE scale is increased. For $m_h=125$~GeV the Higgs quartic crosses zero around $\Lambda=10^9$~GeV and remains at small negative values for higher RGE scales. While this poses a stability problem, are more refined analysis~\cite{EliasMiro:2011aa,Bezrukov:2012sa,Degrassi:2012ry} shows that the SM is stable against vacuum decay on time scales larger than the age of the universe. 

\begin{figure}
\center
\includegraphics{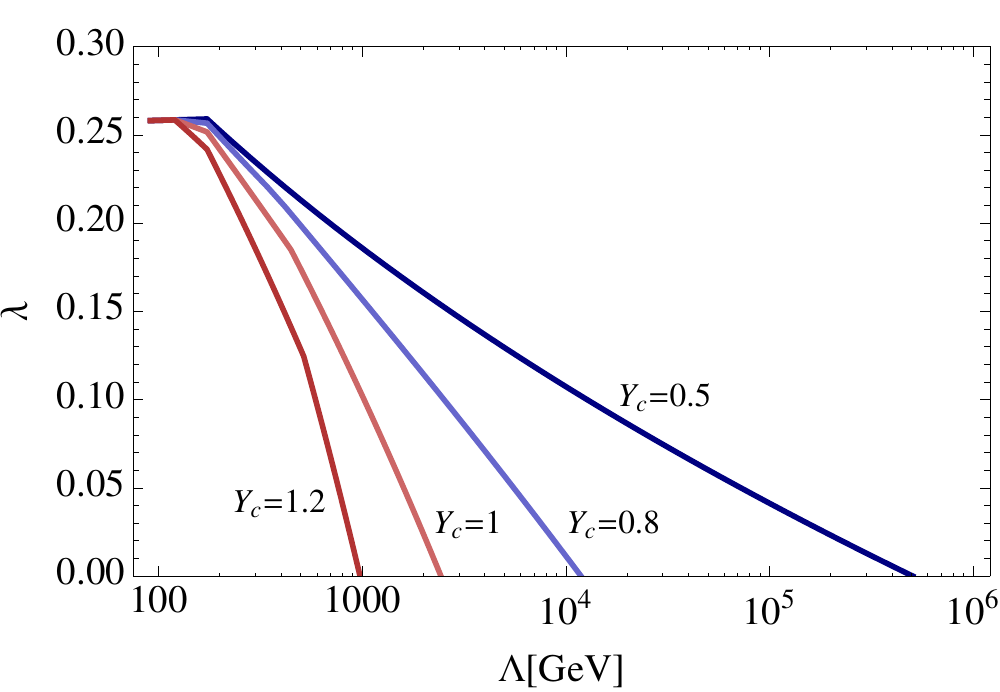}
\caption{Evolution of the Higgs quartic coupling as a function of the scale $\Lambda$, for different values of the charged lepton Yukawa couplings $(Y_c'=Y_c''=Y_c)$, as indicated in the figure. Threshold were taken as $100$~GeV, $173$~GeV for the light charged lepton and top quark, respectively, while the threshold for the heavier charged lepton is given by the relation $M_{E_2} = (2Y_c v + 100$~GeV), provided that $m_\ell = m_e>Y_c v$.  }
 \label{fig:rgg0.8}
\end{figure}

The additional Yukawa couplings in the present model make the model decrease faster. For the numerical study, we assume equal charged lepton Yukawa couplings and vanishing neutrino Yukawas, a parameter choice that is motivated by the results for the Higgs decay rates to photon pairs. The running of $\lambda$ for several choices of $Y_c' = Y_c'' = Y_c$ is shown in Fig.~\ref{fig:rgg0.8}. 

For $Y_c>1$ the running is fast enough to destabilize the Higgs potential at the TeV scale. This would imply the need for a UV completion at or below this scale, which could have sizable contributions to electroweak precision observables and Higgs branching ratios and thus question the validity of our analysis. To avoid this, we restrict our analysis to charged lepton Yukawa couplings smaller than unity. One should further note that small Yukawa couplings for the neutrinos, $Y_n<0.5$, will not have a significant effect on these results, since the destabilizing contributions to the beta function are quartic in the Yukawa couplings. 

The limits on the magnitude of the Yukawa couplings obtained here should be considered conservative. An increased range of validity for the effective theory would be obtained if one allows metastability and demands instead that the vacuum decay time is longer than the age of the universe.

\section{Dark Matter}\label{sec:dm}

So far, we have not considered the mixing of particles of the new sector with SM leptons. Such mixing can be prohibited by making all new leptons introduced in this model odd under a parity symmetry. A natural consequence then is that the lightest particle which is odd under this symmetry becomes  stable on cosmological timescales, and, if it is produced at some point in the early universe, will have a relic density that could be (part of) the dark matter of the universe. 

Leptonic dark matter candidates with unsuppressed couplings to the Z boson, such as ordinary fourth generation neutrinos, are excluded by limits from direct detection, since they predict a large interaction rate with nucleons~\cite{Keung:2011zc}.  Even when a nonzero Majorana mass is introduced to suppress the Z-boson mediated spin independent cross section, the limits on the spin dependent cross section still exclude Majorana neutrino dark matter with masses between $10$~GeV and $2$~TeV~\cite{Angle:2008we,Keung:2011zc,Heikinheimo:2012yd,Jeong:2012fv}. 

This constraint can be relaxed in the model considered here. For vanishing Yukawa couplings, the two singlet neutrinos  $\nu_{\rm R}'$ or $\nu_{\rm L}''$ have no couplings to the Z boson. With nonzero Yukawas the singlet neutrinos will mix with the neutral doublet states and therefore can have nonzero but small couplings to the Z boson. For definiteness, we consider a scenario where only one neutral Yukawa coupling, $Y_n'$, is nonzero, while $Y_n''=m_\nu=0$, so that $\nu_{\rm L}''$ is sterile. The remaining neutrino states mix according to the $3\times 3$ mass matrix
\begin{align}
	{\cal M}_{3 n} & = \begin{pmatrix}
	0 & Y'_n v & m_\ell  \\
	Y'_n v & m' & 0  \\
	m_\ell & 0 & 0 
	\end{pmatrix},
\end{align}
with rows and columns corresponding to $(\nu_{\rm L}',\, \nu_{\rm R}' ,\, \nu_{\rm R}'')$. In the limit where the Yukawa coupling goes to zero, the two neutrinos from the SU(2) doublets form a Dirac neutrino ${\cal N}$ with mass $m_\ell$, while the singlet has a mass $M'$. 

As long as the Yukawa coupling is small, $Y_n' v \ll m_\ell$, and $m' < m_\ell$, 
there will be two heavy neutrino states $N_2$ and $N_3$ with masses $M_{N_{2,3}} \sim m_\ell$ and which are mostly composed of $\nu_{\rm L}'$ and $\nu_{\rm R}''$, while the third neutrino state $N_1$ will have a mass $M_{N_1} \sim m'$ and will be dominantly singlet, with small mixings that generate a nonzero coupling to SM particles that are suppressed by powers of
\begin{align}
	\frac{Y_n'^2 v^2}{m_\ell^2-m'^2}\,.
\end{align}
Motivated by the results of the previous sections, we fix the parameters of the charged lepton sector to $m_\ell=205$~GeV, $m_e=300$~GeV and $Y_c' = Y_c''=0.8$. In this case the lightest charged lepton will be close to the LEP limit, $M_{E_1}=105.8$~GeV, which puts an upper limit on the mass of the dark matter candidate since $M_{\rm DM} \equiv M_{N_1} < M_{E_1}$. 

A suppressed coupling to SM particles also implies that the annihilation rates are suppressed, such that the thermal relic density will be too large for very small Yukawa couplings. It is therefore important to accurately treat the regions where the annihilation is enhanced either resonantly or through co-annihilation effects. For this purpose we have implemented the model into the dark matter code MicrOmegas~\cite{Belanger:2010gh}, and verified the results by comparing with analytic approximations~\cite{Keung:2011zc,Servant:2002aq,Low:2011kp} in regions where they are applicable. The couplings of the charged and neutral lepton mass eigenstates to the SM are given in the appendix. 
\begin{figure}
\center
\includegraphics[width=.48\textwidth]{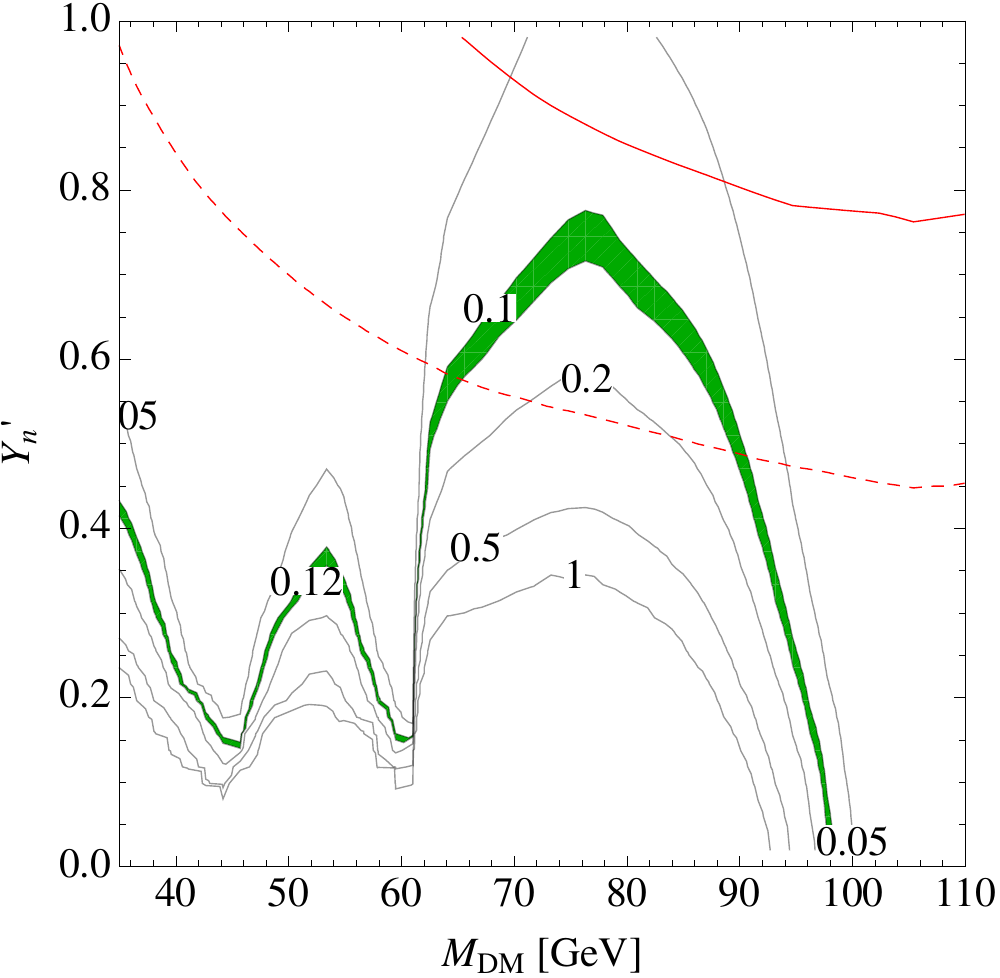}
\includegraphics[width=.48\textwidth]{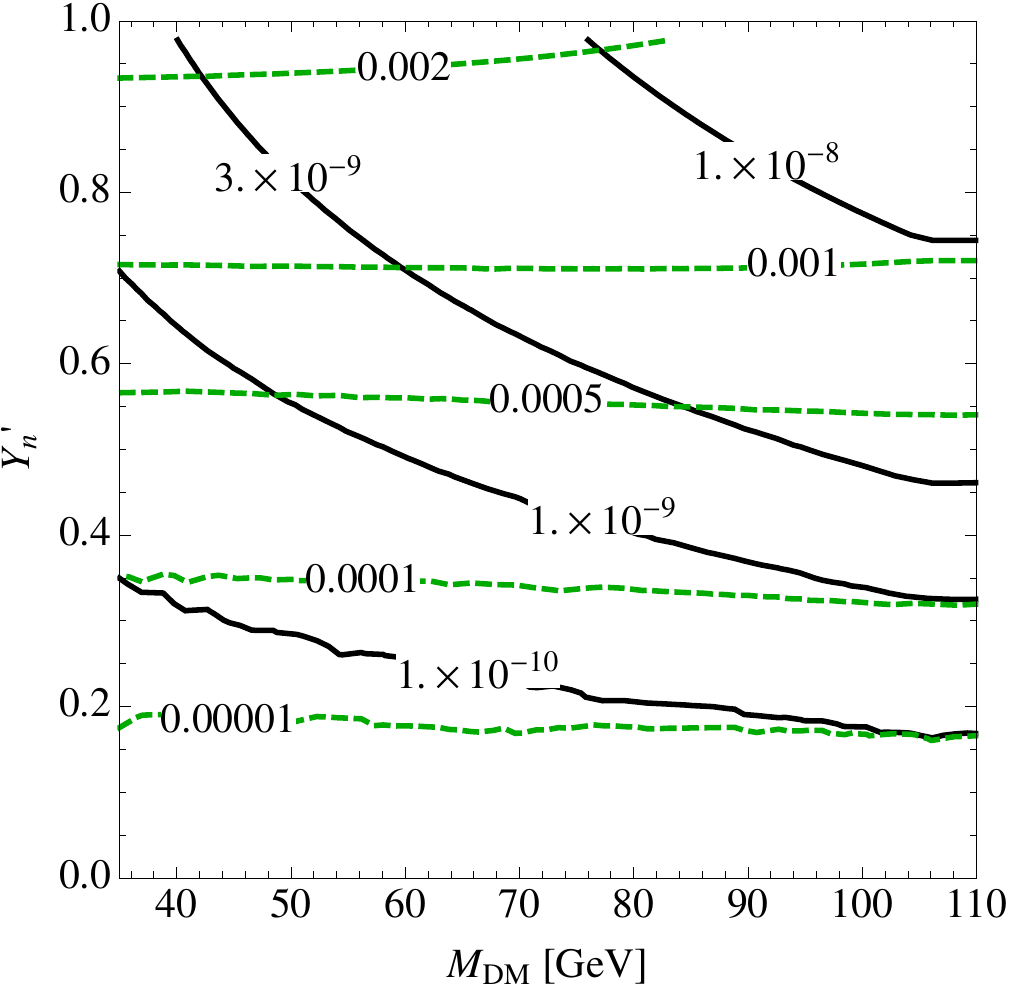}
\caption{Left: Relic density of $\nu_1$ as function of the mass $m_{\nu_1} = m_{\rm DM}$ and of the neutrino Yukawa coupling $Y_n'$. The region where the relic density is consistent with the observed value is shaded green (grey), the shape of the green band is explained in more detail in the text. The red lines indicate the old (solid) and new (dashed) limit on $\sigma_{\rm SI}$ from Xenon~100. Right: SI (black, solid) and SD (green, dashed) cross sections for scattering off protons in pico barn (pb).
}
 \label{fig:dm}
\end{figure}

In Fig.~\ref{fig:dm} we show the relic density as a function of the dark matter mass $m_{\rm DM}$ and the Yukawa coupling $Y_c'$. The two dips at $M_{\rm DM}\sim 45$~GeV and $M_{\rm DM} \sim 62$~GeV are due to resonant annihilation into $Z$ bosons or Higgs bosons respectively. Note that due to the narrowness of the Higgs boson the resonant suppression of the relic density only happens for $M_{\rm DM}$ slightly below $m_h/2$, where the resonant annihilation energy is obtained due to the velocity distribution of the dark matter. Above $M_{\rm DM}=80$~GeV annihilation into $W^+W^-$ pairs becomes kinematically accessible. Finally the strong suppression of the relic density for $M_{\rm DM} \sim 100$~GeV is due to co-annihilation with the lightest charged lepton, which is not velocity suppressed. 

\

As mentioned above, for Majorana particles the spin independent (SI) scattering mediated by Z-bosons is suppressed, such that the most important constraints come from the Higgs mediated SI cross section and from the spin dependent (SD) cross section from Z-boson exchange. For dark matter masses of order $100$~GeV, the strongest limit on the SI cross section, $\sigma_{\rm SI} < (2-3)\times 10^{-45}~{\rm cm}^2 = (2-3) \times 10^{-9}$~pb, comes from the Xenon~100 experiment~\cite{Aprile:2011hi,Aprile:2012}, while the COUPP experiment~\cite{Behnke:2012ys} puts a limit of $\sigma_{\rm SD,p} < 5\times 10^{-39}~{\rm cm}^2 = 5 \times 10^{-3}$~pb on the spin dependent cross section for scattering on protons.

In Fig.~\ref{fig:dm} we show both the SI and SD cross sections for the same parameters as discussed before. While the limits on the SD cross section are not yet imposing a constraint on our model, the updated result of the Xenon~100 experiment~\cite{Aprile:2012} is sensitive to some of the models parameter space, thanks to an improvement over the previous result by about a factor of three. However one should keep in mind that astrophysical uncertainties and variations in the nucleon form factors can easily change these limits by up to a factor of two. 

A strong limit of $\sigma_{SD} < 5\times 10^{-40}~{\rm cm}^2 = 5\times 10^{-4}$~pb was reported by the CMS collaboration~\cite{CMSdm} from searches for mono-photons and missing energy. This limit is not directly applicable here since it assumes that the particles which mediate DM annihilation and scatterings have TeV scale masses. Interpreting these results in our model would require a detailed simulation of the signal in question, which goes beyond the scope of this paper.

In regions where the annihilation rate is not enhanced, Yukawa couplings of order $0.5$ or larger are needed to obtain the correct relic density. This scenario is now strongly constrained by the most recent Xenon~100 results. In the light of the RGE evolution of the Higgs quartic coupling, a smaller Yukawa coupling is also much preferred. The region where $M_{\rm DM} < m_h/2$, where the annihilation rate is enhanced resonantly, will allow the Higgs to decay to pairs of dark matter, leading to a sizable invisible Higgs width, which is disfavored by the most recent LHC results~\cite{Espinosa:2012vu,Giardino:2012dp}.

\begin{figure}
\center
\includegraphics{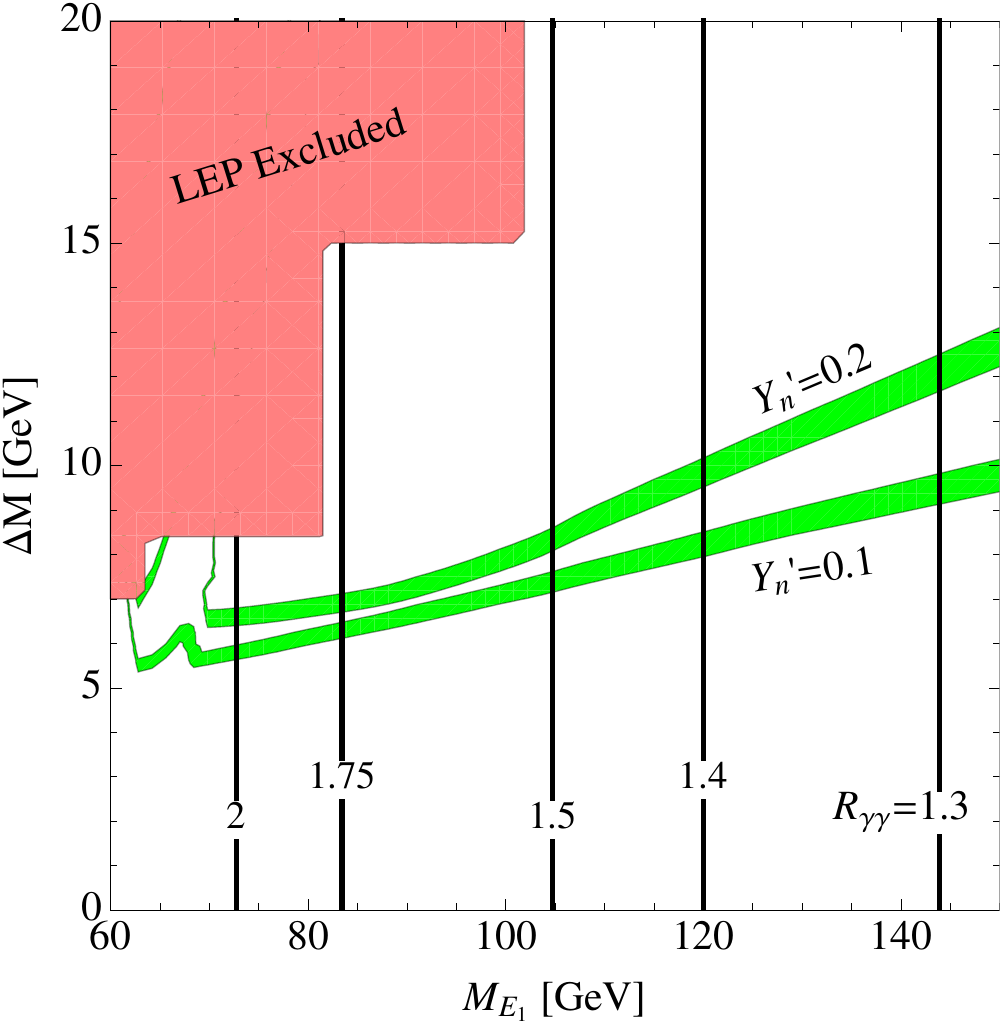}
\caption{LEP limits, relic density and $R_{\gamma \gamma}$ in the co-annihilation regime, as function of the lightest charged lepton mass $m_{e_1}$ and $\Delta m = m_{e_1} - m_{\nu_1}$. The red shaded region is excluded by LEP, the green area is consistent with the relic density constraint for two choices of the neutral Yukawa coupling and the vertical lines are contours of constant $R_{\gamma \gamma}$. See the text for details on the parameter choices. }
 \label{fig:coan}
\end{figure}

The co-annihilation region seems to be most consistent with Higgs data and RGE constraints. Interestingly, this region also allows one to reduce the mass of the lightest charged lepton, since the strongest constraint, $M_{E_1} > 101.9$~GeV only applies when the mass difference $M_{E_1} - M_{N_1}$ is larger than $15$~GeV. We have therefore analyzed this region in more detail, with special attention to the maximally attainable enhancement in the $h \to \gamma \gamma$ rate. We choose $m_\ell = m_e$ and $Y_c' = Y_c''=0.8$ since those choices maximize $R_{\gamma \gamma}$. The mass of the lightest charged lepton is then simply given by 
\begin{align} 
	M_{E_1} = |m_\ell - Y_c v| = |m_\ell - 139.2\text{~GeV} |\,.
\end{align}

The results of our analysis are displayed in Fig.~\ref{fig:coan}, in the $M_{E_1} - \Delta M$ plane, where $\Delta M = M_{E_1} - M_{N_1}$. For a neutral Yukawa coupling of order $0.1$ the preferred mass difference $\Delta M$ is between 5~GeV and 10~GeV, in which case most of the limits on additional charged leptons listed in~\cite{PDG} do not apply. An enhancement of the $h\to \gamma \gamma$ rate of more than 50\% is now easily attainable in regions that are consistent with a thermal relic $N_1$. The regions outside of the green bands can be brought into agreement with the relic density constraint by modifying the neutral Yukawa coupling $Y_n'$. For a fixed coupling, the region below the green bands is allowed if other particles contribute to the dark matter relic density, since $\nu_1$ is over-annihilated in these regions.  
%

\section{Conclusions}\label{sec:conclusions}

In this work, we have presented a compact extension of the SM by a vector-like family of leptons, and shown that it can provide a viable dark matter candidate and a source for the enhancement in the Higgs branching ratio to photon pairs, thus explaining two of the remaining mysteries of the SM. 

We have performed a detailed analysis of electroweak precision constraints, outlining the decoupling effects of the vector-like masses that suppress contributions to the S and T parameters even in the presence of sizable custodial symmetry breaking from Yukawa couplings. It is shown that the Higgs branching ratio can be enhanced if there is mixing in the charged lepton sector, while the case without mixing is excluded at the $3\sigma$ level by the observation of the Higgs boson in the di-photon channel. 

The running of the Higgs quartic coupling places an upper bound on the Yukawa couplings. If one demands that the model is stable up to energies accessible at the LHC, then $Y_c', Y_c'' \lesssim 1$, assuming that the effects of the neutral Yukawas are negligible. This leads to a strong correlation between the maximal attainable value for $R_{\gamma \gamma}$ and the mass of the lightest charged lepton, which has to be  ${\cal O}(100~$GeV) in order to obtain a 50\% enhancement. Such a light charged particle will eventually be in reach of the LHC experiments, thus making the model testable in the near future.

In the absence of mixing between the new lepton sector and the SM leptons, the lightest neutrino is stable. To satisfy the relic density constraint while keeping the neutral Yukawa couplings small, either resonant annihilation or co-annihilation with the lightest charged lepton is needed. The latter possibility is preferred since it also avoids invisible Higgs decays to pairs of dark matter. The preferred mass difference $\Delta M = M_{E_1} - M_{N_1}$ for the co-annihilation scenario is between 5~GeV and 10~GeV. In this regime the strongest LEP limits on $M_{E_1}$ do not apply, and charged lepton masses below $100$~GeV and values of $R_{\gamma \gamma} \sim 2$ become possible. 

At the LHC, the most promising signal will come from Drell-Yan production of $E_1^+ E_1^-$ pairs that subsequently decay through a virtual $W$ boson, $E_1 \to W^{*} N_1$, giving rise to final states with leptons, jets and missing energy. In the co-annihilation regime the leptons and jets from the $W^{*}$ decay become increasingly soft, so that only events where sufficiently hard jets are radiated from the initial state partons will be detectable. In this case better sensitivity can be obtained from $E_1 N_2$ and $N_1 N_2$ production, followed by $N_2 \to W E_1$ or $N_2 \to Z N_1$ decays where now the $Z$ and $W$ bosons can be on-shell. Furthermore monojet and mono-photon + missing energy searches can be employed to place a bound on direct $E_1^+ E_1^-$ and $N_1 N_1$ production rates. These searches might be able to probe the $m_{E_1}\lesssim 120$~GeV regime of our model with a few tens of fb$^{-1}$ in the 14~TeV LHC run~\cite{ArkaniHamed:2012kq}, provided that systematic uncertainties are under control. 

In addition the dark matter candidate can be searched for in direct detection experiments. The SI and SD cross sections are in reach of the next generation of experiments, but the search will be challenging in the co-annihilation regime. 
 
It remains to say that the model is mostly phenomenologically motivated, and should be viewed as an effective theory at the weak scale that will eventually be embedded in a more complete UV theory, which should also ensure vacuum stability beyond the 10~TeV scale. One option would certainly be a supersymmetric completion of the model with a somewhat higher supersymmetry breaking scale and superpartners above the TeV scale. Another intriguing possibility is the completion with a larger number of vector-like families of quarks and leptons that can give successful gauge coupling unification when three complete families are added to the SM~\cite{Dermisek:2012as}.

\acknowledgments{
P.~S. would like to thank the lively atmosphere at the 2012 CERN TH-LPCC summer institute on LHC physics, where part of this work was performed. Work at ANL is supported in part by the U.S. Department of Energy, Division of High Energy Physics, under grant number DE-AC02-06CH11357, and at UIC under grant number DE-FG02-84ER40173. A.~J. acknowledges support from  the Subrahmanyan Chandrasekhar Memorial Fellowship.

\appendix
\allowdisplaybreaks
\section*{Appendix}
\section{Oblique Electroweak Parameters}
Lagrangian for masses of vector-like fourth generation neutrinos is
\begin{align}
	{\cal L_{\nu \cal M}} & = \frac{1}{2} \, \begin{pmatrix}
         \overline{\nu'_{\rm L}} & \overline{{\nu'_\RR}^c} & \overline{{\nu''_\RR}^c} & \overline{\nu''_{\rm L}} \\
        \end{pmatrix}
        \begin{pmatrix}
	0 & Y'_n v & m_\ell & 0 \\
	Y'_n v & m' & 0 & m_\nu\\
        m_\ell & 0 & 0 & Y''_n v\\
        0 & m_\nu & Y''_n v & m''
	\end{pmatrix}\begin{pmatrix}
	{\nu'_\LL}^c\\        
	\nu'_\RR\\ 
        \nu''_\RR\\
        {\nu''_\LL}^c
	\end{pmatrix}+{\rm h.c.}
\end{align}
Yukawa couplings and Majorana masses can be complex in general, making neutrino mass matrix a complex symmetric matrix. Such matrix can be diagonalized using Takagi decomposition as $\mathcal{M}_{n_D}=V^T\mathcal{M}_nV$, where $V$ is a unitary matrix. This gives mixing of left and right chiralities as follows
\begin{align}
	\begin{pmatrix}
	\mathbb{P}_\LL N_1\\        
	\mathbb{P}_\LL N_2\\ 
        \mathbb{P}_\LL N_3\\
        \mathbb{P}_\LL N_4
	\end{pmatrix} = V^T\begin{pmatrix}
	{\nu'_\LL}\\        
	{\nu'_\RR}^c\\ 
        {\nu''_\RR}^c\\
        {\nu''_\LL}
	\end{pmatrix}\quad\quad\quad\text{and}\quad\quad\quad
        \begin{pmatrix}
	\mathbb{P}_\RR N_1\\        
	\mathbb{P}_\RR N_2\\ 
        \mathbb{P}_\RR N_3\\
        \mathbb{P}_\RR N_4
	\end{pmatrix} & = V^\dagger\begin{pmatrix}
	{\nu'_\LL}^c\\        
	\nu'_\RR\\ 
        \nu''_\RR\\
        {\nu''_\LL}^c
	\end{pmatrix}.\label{eq:nudiag}
\end{align}
Let the diagonal entries of $\mathcal{M}_{n_D}$ be $M_{N_{1-4}}$ which are masses of Majorana neutrino states $N_{1-4}$ respectively. The Takagi decomposition produces a positive definite diagonal matrix which ensures the masses are positive.

Lagrangian for masses of fourth generation charged leptons is 
\begin{align}
	{\cal L}_{e \cal M} & = \begin{pmatrix}
         \bar{e}'_\LL & \bar{e}''_\LL\\
         \end{pmatrix}
         \begin{pmatrix}
	Y'_cv & m_\ell \\
	m_e & Y''_cv
	\end{pmatrix}\begin{pmatrix}
        e'_\RR \\ 
        e''_\RR \\
	\end{pmatrix}+h.c.
\end{align}

For chaged lepton states, mass matrix can be diagonalized by the singular value decomposition (SVD) given as $\mathcal{M}_{c_D}=U_L^\dagger\mathcal{M}_cU_R$. The transformation between mass states $E_1,\,E_2$ and flavor and chiral states $e'_\LL,\,e''_\LL,\,e'_\RR,\,e''_\RR$ after singular value decomposition is given as follows:
\begin{align}
        \begin{pmatrix}
	{\mathbb{P}_\LL E_1}\\        
	{\mathbb{P}_\LL E_2}
        \end{pmatrix} & = U_L^\dagger\begin{pmatrix}
	{e'_\LL}\\        
	{e''_\LL}
      	\end{pmatrix}\quad\text{and}\quad
        \begin{pmatrix}
	{\mathbb{P}_\RR E_1}\\        
	{\mathbb{P}_\RR E_2} 
        \end{pmatrix} = U_R^\dagger\begin{pmatrix}
	{e'_\RR}\\        
	{e''_\RR} 
      	\end{pmatrix}.\label{eq:lepdiag}
\end{align}
%
The part of Lagrangian involving fourth generation couplings to $W_\mu^1$ and $W_\mu^3$ boson is given as follows
\begin{align}
{\cal{L}}_{W_\mu^1}&=\frac{g_2}{2}\overline{\nu'_\LL}\gamma^\mu W_\mu^1e'_\LL+\frac{g_2}{2}\overline{e'_\LL}\gamma^\mu W_\mu^1\nu'_\LL+\frac{g_2}{2}\overline{\nu''_\RR}\gamma^\mu W_\mu^1e''_\RR+\frac{g_2}{2}\overline{e''_\RR}\gamma^\mu W_\mu^1\nu''_\RR\,,\label{eq:lagFl1}\\
{\cal{L}}_{W_\mu^3}&=\frac{g_2}{2}\overline{\nu'_\LL}\gamma^\mu W_\mu^3\nu'_\LL-\frac{g_2}{2}\overline{e'_\LL}\gamma^\mu W_\mu^3e'_\LL+\frac{g_2}{2}\overline{\nu''_\RR}\gamma^\mu W_\mu^3\nu''_\RR-\frac{g_2}{2}\overline{e''_\RR}\gamma^\mu W_\mu^3e''_\RR\,.\label{eq:lagFl3}
\end{align}
Substituting~\ref{eq:nudiag},~\ref{eq:lepdiag} in the flavour basis Lagrangian given in~\ref{eq:lagFl1}, we obtain the corresponding Lagrangian in mass basis as follows 
\begin{align}
{\cal L}_{W_\mu^1} &=\frac{g_2}{2} \sum_{j=1}^2\sum_{k=1}^{4}\left(\overline{E_j}\gamma^\mu\frac{U_{L_{1j}}^*V_{1k}^*\mathbb{P}_\LL+U_{R_{2j}}^*V_{3k}\mathbb{P}_\RR}{2}W_\mu^1N_k\right.\notag\\
&\,\,\,\,\quad\quad\quad+\left.\overline{N_k}\gamma^\mu\frac{V_{1k}U_{L_{1j}}\mathbb{P}_\LL+V_{3k}^*U_{R_{2j}}\mathbb{P}_\RR}{2}W_\mu^1E_j\right).
\end{align}
Similarly, we substitute~\ref{eq:nudiag},~\ref{eq:lepdiag} in the flavour basis Lagrangian given in~\ref{eq:lagFl3} to get
\begin{align}
{\cal{L}}_{W_\mu^3}&=g_2\sum_{j,k=1}^4\overline{N_j}\gamma^\mu \frac{V_{1j}V_{1k}^*\mathbb{P}_\LL+V_{3j}^*V_{3k}\mathbb{P}_\RR}{2}W_\mu^3N_k\notag\\
&\,\,\,\,-g_2\sum_{j,k=1}^2\overline{E_j}\gamma^\mu \frac{U_{L_{1j}}^*U_{L_{1k}}\mathbb{P}_\LL+U_{R_{2j}}^*U_{R_{2k}}\mathbb{P}_\RR}{2}W_\mu^3E_k\,.
\end{align}
Part of Lagrangian involving fourth generation couplings to photon $A_\mu$ is given as follows
\begin{align}
{\cal L}_{\gamma_\mu} &=eA_\mu Q\left(\bar{E}_1\gamma^\mu E_1+\bar{E}_2\gamma^\mu E_2\right)=-eA_\mu\left(\bar{E}_1\gamma^\mu E_1+\bar{E}_2\gamma^\mu E_2\right).
\end{align}
From above described lagrangians, we can derive gauge boson self energies $\Pi_{11}\left(q^2\right)$,\,$\Pi_{33}\left(q^2\right)$ and $\Pi_{3Q}\left(q^2\right)$ as given in~\cite{Peskin:1991sw} and use the following formulae to find S and T.
\begin{align}
S & = 16\pi\frac{d}{dq}\left[\Pi_{33}\left(q^2\right)-\Pi_{3Q}\left(q^2\right)\right]\bigg|_{q^2=0},\\
T & = \frac{4\pi}{c_w^2s_w^2M_z^2}\left[\Pi_{11}\left(0\right)-\Pi_{33}\left(0\right)\right].
\end{align}
We define
\begin{align}
\Delta& = M_2^2x+M_1^2\left(1-x\right)-x\left(1-x\right)q^2,\\
b_0\left(M_1,M_2,q^2\right)&=\int_0^1\text{log}\left(\frac{\Delta}{\Lambda^2}\right)\,dx\,,\\
b_1\left(M_1,M_2,q^2\right)&=\int_0^1 x\,\text{log}\left(\frac{\Delta}{\Lambda^2}\right)\,dx\,,\\
b_2\left(M_1,M_2,q^2\right)&=\int_0^1 x\left(1-x\right)\text{log}\left(\frac{\Delta}{\Lambda^2}\right)\,dx=b_2\left(M_2,M_1,q^2\right),\\
b_3\left(M_1,M_2,0\right)&=\frac{M_2^2\,b_1\left(M_1,M_2,0\right)+M_1^2\,b_1\left(M_2,M_1,0\right)}{2}\,,\\
f_3\left(M_1,M_2\right)&=M_1M_2\frac{M_2^4-M_1^4+2M_1^2M_2^2\text{log}\left(M_1^2/\Lambda^2\right)-2M_1^2M_2^2\text{log}\left(M_2^2/\Lambda^2\right)}{2\left(M_1^2-M_2^2\right)^3}\,,
%
\end{align}
where $\Lambda^2$ is an arbitrary regularization scale that will drop out of all physical observables. Note also that $f_3(M_1, M_1) = -1/6$ is well defined in the limit $M_2 \to M_1$.  
Using these we can write\footnote{We thank P.~Grothaus and M.~Fairbairn for pointing out a typographical error in (\ref{eq:T}) in an earlier version of this paper.} 
\begin{align}
\pi S&=\sum_{j,k=1}^2\left(\left|U_{L_{1j}}\right|^2\left|U_{L_{1k}}\right|^2+\left|U_{R_{2j}}\right|^2\left|U_{R_{2k}}\right|^2\right)b_2\left(M_{E_j},M_{E_k},0\right)\notag \\
&\,\,\,\,+\sum_{j,k=1}^2Re\left(U_{L_{1j}}U_{L_{1k}}^*U_{R_{2j}}^*U_{R_{2k}}\right)f_3\left(M_{E_j},M_{E_k}\right)\notag\\
&\,\,\,\,+\sum_{j,k=1}^4 \left(\left|V_{1j}\right|^2\left|V_{1k}\right|^2+\left|V_{3j}\right|^2\left|V_{3k}\right|^2\right)b_2\left(M_{N_j},M_{N_k},0\right)\notag\\
&\,\,\,\,+\sum_{j,k=1}^4\text{Re}\left(V_{1j}V_{1k}^*V_{3j}V_{3k}^*\right)f_3\left(M_{N_j},M_{N_k}\right) \notag \\
&\;\;-2\sum_{j=1}^2\left(\left|U_{L_{1j}}\right|^2+\left|U_{R_{2j}}\right|^2\right)b_2(M_{E_j},M_{E_j},0)
+\frac{1}{3}\,, \label{eq:S} 
\end{align}
\begin{align}
4\pi &s_w^2c_w^2M_z^2T =  \notag
\\
&\;\; -2\sum_{j,k=1}^{2,4}\left(|U_{L_{1j}}|^2\left|V_{1k}\right|^2+|U_{R_{2j}}|^2\left|V_{3k}\right|^2\right)b_3\left(M_{N_k},M_{E_j},0\right)\notag\\
&\,\,\,\,+2\sum_{j,k=1}^{2,4}\,\text{Re}\left(U_{L_{1j}}U_{R_{2j}}^*V_{1k}V_{3k}\right)M_{E_j}M_{N_k}b_0\left(M_{E_j},M_{N_k},0\right)\notag\\
&\,\,\,\,+\sum_{j,k=1}^4 \left(\left|V_{1j}\right|^2\left|V_{1k}\right|^2+\,\left|V_{3j}\right|^2\left|V_{3k}\right|^2\right)b_3\left(M_{N_j},M_{N_k},0\right)\notag\\
&\,\,\,\,-\sum_{j,k=1}^4\,\text{Re}\left(V_{1j}V_{1k}^*V_{3j}V_{3k}^*\right)M_{N_j}M_{N_k}b_0\left(M_{N_j},M_{N_k},0\right)\notag\\
&\,\,\,\,+\left(|U_{L_{11}}|^4+|U_{R_{21}}|^4\right)M_{E_1}^2b_1\left(M_{E_1},M_{E_1},0\right)+\left(|U_{L_{12}}|^4+|U_{R_{22}}|^4\right)M_{E_2}^2b_1\left(M_{E_2},M_{E_2},0\right)\notag\\
&\,\,\,\,+\left(2|U_{L_{11}}|^2|U_{L_{21}}|^2+2|U_{R_{12}}|^2|U_{R_{22}}|^2\right)b_3\left(M_{E_1},M_{E_2},0\right)\notag\\
&\,\,\,\,-\sum_{j,k=1}^2\text{Re}\left(U_{L_{1j}}U_{L_{1k}}^*U_{R_{2j}}^*U_{R_{2k}}\right)M_{E_j}M_{E_k}b_0\left(M_{E_j},M_{E_k},0\right).\label{eq:T}
\end{align}
For the case where the explicit mass terms vanish, i.e. $U_{L_{ij}}=U_{R_{ij}}=\delta_{ij}$ and $V_{jk}\ne0$ only if j and k both belong to either \{1,2\} or \{3,4\},~\ref{eq:S} and~\ref{eq:T} reduce to the following
\begin{align}
\pi S&=\frac{1}{3}-b_2\left(M_{E_1},M_{E_1},0\right)-b_2\left(M_{E_2},M_{E_2},0\right)\notag\\
&+\sum_{j,k=1}^2 \left|V_{1j}\right|^2\left|V_{1k}\right|^2b_2\left(M_{N_j},M_{N_k},0\right)+\sum_{j,k=3}^4 \left|V_{3j}\right|^2\left|V_{3k}\right|^2 b_2\left(M_{N_j},M_{N_k},0\right),\label{eq:SI}
\end{align}
\begin{align}
4\pi s_w^2c_w^2M_z^2T&=
\sum_{j=1}^2 M_{E_j}^2 b_1(M_{E_j},M_{E_j},0)
-2\sum_{j=1}^2\,\left|V_{1j}\right|^2b_3\left(M_{N_j},M_{E_1},0\right)\notag\\
&\,\,\,\,-2\sum_{j=3}^4\,\left|V_{3j}\right|^2b_3\left(M_{N_j},M_{E_2},0\right)+\sum_{j,k=1}^2\left|V_{1j}\right|^2\left|V_{1k}\right|^2b_3\left(M_{N_j},M_{N_k},0\right)\notag\\
&\,\,\,\,+\sum_{j,k=3}^4\,\left|V_{3j}\right|^2\left|V_{3k}\right|^2b_3\left(M_{N_j},M_{N_k},0\right)\,.\label{eq:TI}
\end{align}
~\ref{eq:SI} and~\ref{eq:TI} in turn reduce to the formulas given in~\cite{Eberhardt:2010bm} in the limit of vanishing Majorana masses. In the decoupling limit where all Yukawas and Majoranas go to zero, we get $S=T=0$ as expected for a vector-like generation as shown in Fig.~\ref{fig:st2}.
             

\end{document}